\documentclass[10pt,twocolumn,letterpaper]{article}

\usepackage{times}
\usepackage{epsfig}
\usepackage{graphicx}
\usepackage{amsmath}
\usepackage{amssymb}
\usepackage{enumitem}
\usepackage{grffile}
\usepackage{tabularx}
\newcommand\setrow[1]{\gdef\rowmac{#1}#1\ignorespaces}
\newcommand\clearrow{\global\let\rowmac\relax}
\clearrow

\graphicspath{{./figures/}}

\usepackage[abs]{overpic}  
\newcommand{\overimg}[3][]{%
    \begin{overpic}[#1]{#2}%
      \put (75, 2) {%
        \setlength{\fboxsep}{2pt}%
        \colorbox{white}{%
          \scriptsize\sffamily\vphantom{y}%
          #3%
        }%
      }%
    \end{overpic}%
}

\newcommand{\overimgx}[3][]{%
    \begin{overpic}[#1]{#2}%
      \put (2, 2) {%
        \setlength{\fboxsep}{2pt}%
        \colorbox{white}{%
          \scriptsize\sffamily\vphantom{y}%
          #3%
        }%
      }%
    \end{overpic}%
}

\usepackage[table]{xcolor}    
\usepackage{array,multirow}
\usepackage{soul}
\usepackage{floatrow}
\newfloatcommand{capbtabbox}{table}[][\FBwidth]
\newcolumntype{H}{>{\setbox0=\hbox\bgroup}c<{\egroup}@{}}
\usepackage{booktabs}
\newcommand{\gf}[2]{{}{\protect#2}}
\newcommand{\nada}[1]{}

\newcommand{\A}{DN}
\newcommand{\B}{DM}

\usepackage[sort&compress, square, numbers]{natbib}

\usepackage[pagebackref=true,breaklinks=true,letterpaper=true,colorlinks,bookmarks=false]{hyperref}

\begin{document}

\title{A Review of an Old Dilemma:  \\
Demosaicking First, or Denoising First? }

\author{Qiyu Jin\\
School of mathematical science, \\
Inner Mongolia University\\ 
{\tt\small qyjin2015@aliyun.com}
\and
Gabriele Facciolo  \\
Centre  Borelli, \\
ENS Paris-Saclay, CNRS\\
{\tt\small  facciolo@cmla.ens-cachan.fr}
\and
Jean-Michel Morel\\
Centre  Borelli,\\
ENS Paris-Saclay, CNRS\\
{\tt\small moreljeanmichel@gmail.com}
}

\maketitle
\thispagestyle{empty}

\begin{abstract}
Image denoising and demosaicking are the most important early stages in digital camera
pipelines. They  constitute a severely ill-posed problem that aims at reconstructing a full
color image from a noisy color filter array (CFA) image.
In most of the literature, denoising and demosaicking are treated as two independent
problems, without considering  their interaction, or asking  which should be applied first.
  Several recent works have started addressing them jointly in  works that involve  heavy
  weight CNNs, thus incompatible with low power portable imaging devices. Hence, the
  question of how to combine denoising and demosaicking to reconstruct full color images
  remains very relevant:
Is  denoising to be applied first, or should that be  demosaicking first?
In this paper, we review  the main variants of these strategies and carry-out an extensive
 evaluation to find the best way to reconstruct full color images from a noisy mosaic.
 We conclude that demosaicking should applied first, followed by  denoising.
 Yet we prove that  this requires an adaptation of classic denoising algorithms to
 demosaicked noise, which we justify and specify.

\end{abstract}

\section{Introduction}

Most digital cameras capture image data by using a single sensor coupled with a color filter array (CFA).
At each pixel in the array, only one color component  is recorded and the resulting image is called a mosaic.
The most common  CFA is the Bayer color array \cite{bayer1976color}, in which
two out of four pixels measure the green component, one measures the red and one the blue.  
The process of completing the missing red, green and blue values at each pixel is called demosaicking.
Noise is inevitable, especially in low light conditions and for small camera sensors like those used in mobile phones.
The conventional approach in image restoration pipelines for processing  noisy raw sensor data has long been to
apply denoising and demosaicking as two independent steps ~\cite{paliy2007demosaicing}. Furthermore,
the immense majority of image processing  papers addressing  one  of  both operations  do not address its
combination with the other one. All classic denoising  algorithms have been designed for color or grey level images
with  white noise added.  Yet the  realistic data are different: either a mosaic with white noise or a demosaicked
image with structured noise.

\paragraph{Joint denoising/demosaicking methods.}

This has led several recent works to propose joint demosaicking- denoising methods~\cite{hirakawa2006joint,khashabi2014joint,chatterjee2011noise,gharbi2016deep}.
 For  example
\cite{goossens2015overview}  proposed a variational model to solve jointly demosaicking, denoising and deblurring. It uses a sparsifying prior based on  wavelet packets and applied on decorrelated color channels. More detail about the  technicalities of this sophisticated method can be found in  \cite{aelterman2013computationally}.
Life has become far easier for joint denoising/demosaicking with the emergence of
 machine learning methods.  It is, indeed, easy to simulate as many learning data as needed. This  methodology can be used to  obtain ground breaking  demosaicking algorithms such as \cite{syu2018learning}. This paper proposed in 2018 a demosaicking CNN outperforming the best handcrafted algorithms including ARI \cite{monno2017adaptive} by nearly 2 decibels.
In \cite{khashabi2014joint}   a public ground truth  dataset was introduced and used for  one of the first joint  demosaicking and denoising methods based on machine learning.  In a rapid succession,  two state of the  art  denoising+demosaicking methods  based on deep learning were proposed: \cite{gharbi2016deep} and \cite{klatzer2016learning}. This last paper performs joint denoising and demosaicking  by a customized neural network presented as  a cascade of energy minimization methods tuned by learning. The outstanding  results beat the claimed anterior best method \cite{Heide2014FlexISP} by 1 decibel. Then in 2018  we have two still better performing methods,  \cite{dong2018joint} involving a GAN, which compares favorably to  \cite{gharbi2016deep} and \cite{tan2017color}. The  method recently proposed in~\cite{kokkinos2019iterative} performs joint denoising and demosaicking by inserting many residual denoising layers in a  CNN. This complex method is claimed to beat~\cite{gharbi2016deep} and~\cite{klatzer2016learning} by a good margin.
Lastly, in  2019, \cite{ehret2019joint}  introduced a ``mosaic-to mosaic'' training strategy analog to the noise-to-noise~\cite{Lehtinen2018}
 and frame-to-frame~\cite{Ehret2019model} frameworks to handle noisy mosaicked raw data,  and training both demosaicking and joint denoising and demosaicking networks without requiring ground truth.  The method starts from pairs or bursts of raw images of the same scene.  It registers them and learns to predict the missing colors.

Yet the question of how to combine  denoising and demosaicking algorithms conceived as independent  blocks remains very relevant, especially in the context of low power or portable devices,  and given the  fact that the main effort in denoising and demosaicking has addressed them independently.
A big argument  in favour of performing  denoising before demosaicking  is that most existing demosaicking algorithms
have  been developed under the  unrealistic assumption of noise-free
data~\cite{hamilton1997adaptive,kiku2013residual,he2013guided,pekkucuksen2010gradient,kiku2014minimized,kiku2016beyond,jaiswal2014exploitation,wu2016demosaicing,monno2017adaptive,kim2016four,chen2015multidirectional,wang2015bayer,buades2011self,
zhang2011color,mairal2009non,gharbi2016deep,tan2017color,tan2018deepdemosaicking,kokkinos2019iterative}.
Yet the performance of these algorithms can degrade
dramatically when the noise level increases on the CFA raw image.
Therefore, a previous denoising step is implicitly required by these algorithms.

\gf{In this paper we take the usual simplifying assumption that the noise in the raw mosaic is additive white Gaussian (AWGN) and that the noise level is known.  This  is realistic because,  first, an Anscombe transform~\cite{Anscombe1948} applied to a raw image results in a nearly AWG noise  and, second, because many accurate  noise evaluation methods exist.}{
In this paper we focus on the early CFA processing in the imaging pipeline (operating in linear space). We assume that the noise in the raw mosaic is additive white Gaussian (AWGN) and that its variance is known.
This is realistic because, first, a \emph{variance stabilizing transform} (VST)~\cite{Anscombe1948} applied to a raw image results in a nearly AWG noise and, second, because an accurate noise model is often known  or can be estimated~\cite{ponomarenko2007automatic,ponomarenko_ipol2013}.}
In general, image denoising
methods can be grouped into two major categories, the model based
methods such as non-local means~\cite{buades2005review,jin2018convergence,jin2017nonlocal}, nlBayes~\cite{lebrun2013nonlocal}, CBM3D~\cite{dabov2006image} and WNNM~\cite{gu2014weighted}, and  deep learning methods
 such as~\cite{jain2009natural,zhang2017beyond}.
The ensuing CNNs are sometimes
flexible in handling denoising problems with various noise
levels.

Our goal here is to determine which strategy is
more advantageous for coupling demosaicking and denoising : Is it applying denoising and then demosaicking (which we will denote $\A\&\B$: $\A$ and  $\B$ indicate denoising  and demosaicking  respectively), or is it better to apply  first demosaicking and then denoising ($\B\&\A$)?

\paragraph{\A\&\B~methods (\emph{i.e.} denoising then demosaicking): advantages and drawbacks.}
Many state of  the art works \cite{paliy2007demosaicing,park2009case,kalevo2002noise,zhang2009pca}
support the opinion that
$\A\&\B$ outperforms $\B\&\A$.
Their first convincing argument is that  after demosaicking noise becomes correlated, thus losing its independent identically distributed (i.i.d.)  white Gaussian property.  This increases the difficulty of applying efficient denoising and actually seems to discard all classic algorithms, that mostly rely on  the AGWN  assumption. A second obvious argument is that the best demosaicking algorithms have been designed with noise-free images.

For example, Park \emph{et al.}~\cite{park2009case} considered the classic Hamilton-Adams (HA) \cite{hamilton1997adaptive} and  \cite{dubois2005frequency} for demosaicking, combined with two denoising methods, BLS-GSM~\cite{portilla2003image} and CBM3D~\cite{dabov2007color}. 
This combination  raises the  question of adapting CBM3D to a CFA.   To do so, the  authors apply a sparsifying 4D color transform \nada{of the (Y, U, V) type,}{to the 4-channel image formed by rearranging the Bayer pixels,} apply BM3D to each channel, then apply the inverse color transform.
In the very  same vein, in the BM3D-CFA method~\cite{danielyan2009cross}  BM3D is applied directly on the CFA color array. To do so, ``only blocks having the same CFA configuration are being compared to build the 3D blocks.
 This is the only modification  of  the original BM3D''.  A little thought leads to the conclusion that this amounts to  denoise four different mosaics of  the same  image  before aggregating the four values obtained  for each  pixel.   The authors compare two denoising algorithms with two different setups: a) filtering CFA as a single image and b) splitting the CFA into four color components, filtering them separately, and recombining back the denoised CFA image. This paper showed a systematic improvement over~\cite{zhang2009pca}.  They use Zhang-Wu \cite{zhang2005color} as demosaicking method for their comparison of result after demosaicking.
 In our comparisons the method of~\cite{danielyan2009cross} will be mentioned every time we consider the $\A\&\B$ setup with BM3D. We will  nevertheless replace 
the  demosaicking of~\cite{zhang2005color} by RCNN~\cite{tan2017color} or RI~\cite{kiku2013residual}, which clearly outperform it. 

Similarly in~\cite{chatterjee2011noise} denoising is performed by  an adaptation of NL-means to the Bayer pattern, where only patches with the same CFA configuration are being matched.  This  paper formulates the demosaicking as a super-resolution problem, assuming that the observed values are actually averages of four values in the high resolution image.  It  then guides this super-resolution problem by the NL-means weights.  The method is compared with \cite{menon2009joint} and  \cite{zhang2009pca}.
The authors of \cite{zhang2009pca} also propose an $\A\&\B$ method,  where  the demosaicking method  is \cite{zhang2005color}  and the
the  denoising method is an adaptation of nlBayes \cite{lebrun2013nonlocal} to a Bayer pattern. First, the method extracts blocks with similar configuration in the  Bayer array and groups them by similarity, then it applies to them PCA and a Wiener denoising procedure which can be also interpreted as an LMMSE.
In our experiments, 
{ 
this PCA method~\cite{zhang2009pca} will be considered every time we evaluate the $\A\&\B$ scheme (but combined with a more recent  demosaicking algorithm such as RCNN~\cite{tan2017color}).}
 The more recent paper \cite{zhang2014joint} involves similar arguments.
This  paper uses \cite{alleysson2005linear}, a linear filter to extract  the luminance from the CFA. Then it remarks that this luminance is correlated, so it applies a variant of NL-means that attempts to decorrelate the noise. The same method is applied to each downsampled color channel and the high frequency of the grey level is transported back to the color channels.   This method under-performs with respect to others considered here, so we shall not include it to our final comparison tables. Nevertheless, it  remains of interest as a fast method compatible with low power cameras.  The paper proves that it has a performance very close to a combination of \cite{zhang2009pca} and \cite{hirakawa2006joint}.

  The paper \cite{patil2016poisson} is another method promoting denoising before demosaicking,  involving dictionary learning methods to remove the Poisson
noise from the single channel images prior to demosaicing. Experimental results on
simulated noisy images as well as real camera acquisitions, show the advantage of these
methods over approaches that remove noise subsequent to demosaicing.  The paper nevertheless uses~\cite{malvar2004high} which is a  historic but outdated demosaicking method.

To summarize, in the  $\A\&\B$ strategy all classic denoising  algorithms such as CBM3D,  nlBayes, nlMeans have been adapted to handle a noisy mosaic where only one of R, G or B is known at each pixel.  Several of them \cite{paliy2007demosaicing,park2009case,kalevo2002noise,zhang2009pca} address this realistic case  by processing the noisy CFA images as a  half-size 4-channel color image (with one red, two green and one blue channels) and  then apply a multichannel denoising algorithm to it.  The advantage of the denoising step of $\A\&\B$ is that the Poisson noise can be led back by the classic Anscombe transform to the  case of i.i.d.  white Gaussian,  and the disadvantage is that the resolution of the image is reduced and, as a result, some details might be lost after denoising. Another issue of this strategy is that the spatial relative positions of the R, G, and  B pixels are lost by handling the image as a four channel half size image.

In this paper, we address the above mentioned  issues.  We shall first delve into the advantages and disadvantages of  $\A\&\B$ and $\B\&\A$ approaches.
We shall then analyze noise properties after demosaicking and adjust two existing classic denoising algorithms (CBM3D and nlBayes) to accommodate them to this type of noise. Then,  we shall perform a thorough experimental evaluation that will lead us to conclude  that  $\B\&\A$ (with an adjusted noise parameter) is superior to $\A\&\B$. This result is opposite to the conclusion of \cite{paliy2007demosaicing,park2009case,kalevo2002noise,zhang2009pca}. The advantages of $\B\&\A$ seem to be linked to the fact that this scheme does not handle a half size 4-channels color image; it therefore uses the classic denoising methods directly on a full resolution color image;  this results in more details being preserved and avoids checkerboard effects.

The rest of the paper is structured as follows. In Section~\ref{sec:framework} we present in detail the problem and the main ideas behind the proposed demosaicking and denoising strategy.  Section~\ref{sec:experiment} is a detailed evaluation of the proposed strategy. Section~\ref{sec:conclusion} concludes.


\section{The demosaicking and denoising  framework}
\label{sec:framework}

\begin{figure}
\begin{center}
\renewcommand{\arraystretch}{0.5} \addtolength{\tabcolsep}{-5pt} \vskip3mm {%
\includegraphics[width=0.35\linewidth]{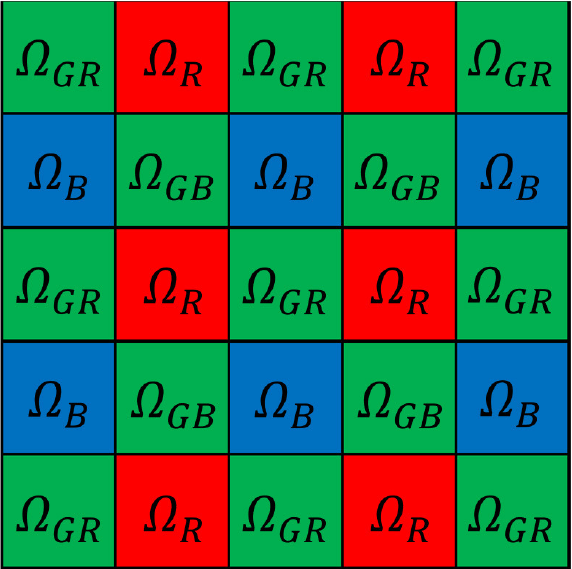}
}
\vspace{-.8em}
\caption{\protect  {Bayer color filter array, CFA, which is  used by most  cameras.}}
\label{Fig bayerfra}
\end{center}
\end{figure}

In a single-sensor camera equipped with a color filter array (CFA) \cite{bayer1976color}, only one pixel value
among the three RGB values is recorded at each pixel. Consider a CFA block as shown in Fig.~\ref{Fig bayerfra}. The raw Bayer CFA images are scalar mosaic matrices with noise.  Obtaining high quality color images requires completing the missing color channels and removing the noise. As mentioned in the introduction, for this task we will consider  two main schemes: $\B\&\A$ (demosaicking then denoising) and  $\A\&\B$ (denoising then demosaicking).

Park {\em et al.}~\cite{park2009case} argued that  demosaicking
introduces chromatic and spatial correlations to the noisy CFA image. Then the noise is no longer i.i.d. white Gaussian, which makes it  harder to remove.
In~\cite{kalevo2002noise},
{some experiments were done to show that  $\A\&\B$ schemes are more efficient to suppress noise  than $\B\&\A$ schemes.}
Based on this argument several denoising methods~\cite{park2009case,zhang2009pca,akiyama2015pseudo,lee2017denoising} for raw CFA images before demosaicking were introduced. Other denoising methods that are not explicitly designed to handle raw
 CFA images (such as  CBM3D and nlBayes) can also be applied  on noisy CFA images by rearranging
the CFA image into a half-size four-channels image
{with  two greens on which the denoising algorithm is applied~\cite{park2009case}.} The denoised CFA is then recovered by undoing the pixel rearrangement.
However, this strategy reduces the resolution of the image seen by the denoiser, and we observed checkerboard effects resulting from chromatic aberrations in the two green channels after denoising. 
To address this issue, Danielyan \emph{et al.}~\cite{danielyan2009cross} proposed BM3D-CFA which amounts to  denoise four different mosaics of  the same  image  before aggregating the four values obtained for each  pixel.

\begin{figure}[t!]
\begin{center}
\renewcommand{\arraystretch}{0.2} \addtolength{\tabcolsep}{-5pt} \vskip3mm {%
\small
\begin{tabular}{llllll}
\multicolumn{2}{c}{\includegraphics[width=0.23\linewidth]{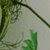}}&\multicolumn{2}{c}{\includegraphics[width=0.23\linewidth]{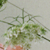}}\\
\multicolumn{2}{c}{(a1) Ground truth }&\multicolumn{2}{c}{(a2) Ground truth }
\\
\includegraphics[width=0.23\linewidth]{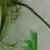}&
\includegraphics[width=0.23\linewidth]{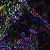}&
\includegraphics[width=0.23\linewidth]{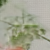}&
\includegraphics[width=0.23\linewidth]{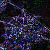}\\
\multicolumn{2}{c}{(b1)${\A}\&\B$  $/26.92$dB}&\multicolumn{2}{c}{(b2)${\A}\&\B$  $/26.92$dB}\\
\includegraphics[width=0.23\linewidth]{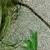}&
\includegraphics[width=0.23\linewidth]{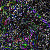}&
\includegraphics[width=0.23\linewidth]{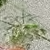}&
\includegraphics[width=0.23\linewidth]{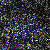}\\
\multicolumn{2}{c}{(c1) $\B\&{\A}$ $/25.38$dB}& \multicolumn{2}{c}{(c2) $\B\&{\A}$ $/25.38$dB}\\
\includegraphics[width=0.23\linewidth]{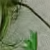}&
\includegraphics[width=0.23\linewidth]{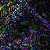}&
\includegraphics[width=0.23\linewidth]{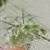}&
\includegraphics[width=0.23\linewidth]{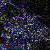}\\
\multicolumn{2}{c}{(d1) $\B\&1.5 {A}$  $/26.95$dB}&\multicolumn{2}{c}{(d2) $\B\&1.5 {A}$  $/26.95$dB}\\
\includegraphics[width=0.23\linewidth]{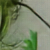}&
\includegraphics[width=0.23\linewidth]{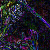}&
\includegraphics[width=0.23\linewidth]{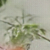}&
\includegraphics[width=0.23\linewidth]{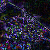}\\
\multicolumn{2}{c}{(e1) JCNN  $/27.46$dB}&\multicolumn{2}{c}{(e2) JCNN  $/27.46$dB} \\
\end{tabular}
} \vskip1mm
\par
\rule{0pt}{-0.2pt}%
\par
\vskip1mm 
\end{center}
\vspace{-.8em}

\caption{{\protect Comparison  of different denoising and demosaicking schemes with noise  $\sigma_0 = 20$. In each experiment: left, detail of the demosaicked and denoised image; right, the difference with original that should contain mainly noise.   $\A$: CBM3D denoising; 
$\B$: demosaicking (here we use RCNN). $1.5\A$ means that if noise level is $\sigma_0$,  the input noise level parameter of denoising method $\A$ is $\sigma=1.5\sigma_0$; $\A\&\B$: uses the BM3D-CFA framework~\cite{danielyan2009cross} for denoising. }}
\label{Fig: bookp1}
\end{figure}

\nada{In the case of splitting into two half-size 3-channel images, both images are  denoised independently and then the pixels are recombined. Each half-size image contributes one green pixel to the denoised CFA image, while the red and blue pixels are averaged. As before, we observed that due to the rearrangement of the CFA pixels many image details are lost in the  image after applying this $\A\&\B$ scheme. 
In addition we observed that this procedure introduces serious checkerboard effects especially visible when the noise level is significant, e.g. $\sigma_0>10$. This is due to  chromatic aberrations in the two parts of the green image.
These artifacts can be observed in  Fig.~\ref{Fig: bookp1} 
(b1)  and 
 (b2).}{}


\paragraph{Modeling demosaicking noise.}
In order to solve the above two problems,
we shall revisit
the $\B\&\A$ scheme.
Compared to the $\A\&\B$ scheme, the advantage of  $\B\&\A$  is that it does not halve the image size. This is a  way around the above mentioned problems. A serious drawback,  though, is that chromatic and spatial correlations have been introduced by the demosaicking in the noise of the CFA image. The result is that the noise is no longer white.
We next analyze some properties of the \textit{demosaicked noise}.

\paragraph{\textbf Definition} \textit{
Given a ground truth color image $(R,G,B)$ we define the {\rm demosaicked noise} associated with a demosaicking method $\B$ in the following  way: first the image is  mosaicked so that  only one value of either $R,G,\B$ is kept at  each pixel, according to a fixed Bayer  pattern. Then white  noise with standard deviation $\sigma_0$ is added to the mosaicked image, and  the resulting noisy mosaic is demosaicked by $\B$, hence giving a noisy image $(\tilde R, \tilde G, \tilde B)$. We call demosaicked noise the  difference $(\tilde R-R, \tilde G-G, \tilde  B-B)$.  In short, it is the
 difference between the demosaicked version of a noisy image and its underlying ground  truth. }

 \medskip

 The  model of the  demosaicked noise  depends on the choice of the demosaicking  algorithm $\B$. For the demosaicking step we will evaluate the following state of  the art methods, which have an increasing complexity:   HA~\cite{hamilton1997adaptive},   RI~\cite{kiku2013residual}, MLRI~\cite{kiku2014minimized},  ARI~\cite{monno2017adaptive}, LSSC~\cite{mairal2009non}, RCNN~\cite{tan2017color} and JCNN~\cite{gharbi2016deep}. We are   interested in algorithms with low  or  moderate power; only HA, RI, MLRI and RCNN have a reasonable complexity in this context.
For the denoising step we shall likewise consider two classic hand-crafted algorithms, CBM3D and nlBayes.

Fig.~\ref{Fig: bookp1}
 (c1) and 
  (c2) shows an example where noisy CFA  images with noise of standard deviation $\sigma_0$ were first  demosaicked by  RCNN and then   restored by
  CBM3D assuming a noise parameter $\sigma = \sigma_0$.  The output of CBM3D with
  $\sigma=\sigma_0$ has a strong residual noise. 
Similar results are also obtained with nlBayes (see the supplementary material).
To understand empirically the  right
  noise model to adopt after demosaicking, we simulated this $\B\&\A$ pipeline for
  different levels of noise $\sigma_0$, and applied  CBM3D after demosaicking
  with a noise parameter corresponding to  $\sigma_0$ multiplied by different
  factors $(1.0,1.1, \cdots, 1.9)$. 

\begin{table}
\caption{
{Denoising after demosaicking $\B\&\A$, 
where $\A$ is CBM3D~\cite{dabov2007color}  with noise parameter equal to $C\, \sigma_0$, while noise in the raw image has standard deviation $\sigma_0=20$.
Each row shows the CPSNR result for  $C$ ranging from $1.0$ to $1.9$. Each column corresponds to a different  demosaicking method $\B$. The best result of each column  is marked with a \fbox{box}.  The best result of each line is in \textcolor{red}{\bf red} and the second best one is in \textcolor{green}{\bf green}.  The best factor $C$ for all methods is $C\simeq 1.5$, the same is true  for different values of $\sigma_0$ as well (see supplementary material).
} }
\vspace{-.8em}
\label{Table bm3d}
\begin{center}
\renewcommand{\arraystretch}{1.25} \addtolength{\tabcolsep}{-3.5pt} \vskip3mm
\small

\begin{tabular}{H c| c c c c H c c c H}
      \rowcolor{gray!50}

\hline
$\sigma_0$ & $C$ &HA&GBTF&RI&MLRI&WMLRI&ARI&LSSC&RCNN&JCNN\\
\hline
  & 1.0 &28.15& 27.58& 28.46& 27.95& 27.83& 28.70& 27.19& 27.28& 27.89\\
  & 1.1 &28.56& 28.15& 28.83& 28.44& 28.35& 28.98& 27.89& 28.05& 28.53\\
  & 1.2 &28.85& 28.55& 29.08& 28.80& 28.74& 29.18& 28.43& 28.67& 29.03\\
  & 1.3 &29.05& 28.81& 29.23& 29.03& 28.99& 29.29& 28.78& 29.09& 29.36\\
  & 1.4 &29.18& 28.96& \textcolor{green}{\bf 29.31}& 29.17& 29.15& \textcolor{red}{\bf        29.35}& 29.00& 29.34& \textcolor{blue}{\bf 29.54}\\
20& 1.5 &\textcolor{blue}{\bf29.23}& \textcolor{green}{\bf 29.00}& \textcolor{red}{\bf 29.32}& \textcolor{green}{\bf 29.22}& \textcolor{green}{\bf 29.21}& \textcolor{red}{\bf 29.35}& \textcolor{red}{\bf 29.06}& \textcolor{red}{\bf 29.41}& \textcolor{red}{\bf 29.58}\\
  & 1.6 &\textcolor{red}{\bf 29.25}& \textcolor{red}{\bf 29.01}& \textcolor{blue}{\bf29.30}& \textcolor{red}{\bf 29.23}& \textcolor{red}{\bf 29.22}& \textcolor{blue}{\bf 29.33}& \textcolor{red}{\bf 29.06}& \textcolor{red}{\bf 29.41}& \textcolor{green}{\bf 29.57}\\
  & 1.7 &\textcolor{red}{\bf 29.25}& \textcolor{blue}{\bf28.97}& 29.26& \textcolor{blue}{\bf29.20}& \textcolor{green}{\bf 29.20}& 29.29&  \textcolor{blue}{\bf 29.02}&  \textcolor{blue}{\bf 29.36}& 29.51\\
  & 1.8 &29.22& 28.92& 29.20& 29.15& 29.15& 29.23& 28.95& 29.28& 29.43\\
  & 1.9 &29.17& 28.85& 29.13& 29.08& 29.08& 29.17& 28.88& 29.20& 29.34\\\hline

\end{tabular}
\end{center}
\end{table}

\begin{table}
\caption{RMSE between
 original and demosaicked image for different demosaicking algorithms  in presence of noise of standard deviation $\sigma_0$. 
 }
\label{Table rmse}
\vspace{-.8em}
\begin{center}
\renewcommand{\arraystretch}{1.25} \addtolength{\tabcolsep}{-2.5pt} \vskip3mm
\small
\rowcolors{2}{gray!25}{white}
\begin{tabular}{c|ccccHcccH}
      \rowcolor{gray!50}

\hline
$\sigma_0$&HA&GBTF&RI&MLRI&MLRI&ARI&LSSC&RCNN&JCNN\\
 \hline
$1$&5.04&  5.10&  4.17&  4.06&  4.03&  3.72&  4.40&  3.21&  3.14\\
$5$&6.78&  6.87&  6.12&  6.10&  6.08&  5.74&  6.36&  5.59&  5.45\\
$10$&10.18& 10.27&  9.53&  9.74&  9.76&  9.09&  9.96&  9.65&  9.18\\
$20$&17.75& 17.83& 16.77& 17.56& 17.66& 16.06& 18.16& 18.04& 17.08\\
$40$&32.67& 32.76& 30.77& 32.64& 32.90& 29.36& 33.68& 33.98& 31.28\\
$60$&46.14& 46.35& 43.43& 46.11& 46.49& 41.44& 48.11& 47.95& 42.57\\
\hline
\end{tabular}
\end{center}
\end{table}

The results are shown in Table~\ref{Table bm3d},
where the classic color peak signal-to-noise ratio (CPSNR) \cite{alleysson2005linear} is adopted as a logarithmic measure of the performance of the algorithms.  
It  is defined by
\begin{align*}
      \textstyle  \mathrm{CPSNR} = 10 \log_{10} \frac{255^2}{\sum_{X=R,G,B} \mathrm{MSE} (X)/3}, 
      \quad \text{with}\\
  \textstyle  \mathrm{MSE}(X) = \frac{1}{|\Omega|} \sum_{ (i,j)\in \Omega} ( \widehat{X}(i,j) -  X(i,j))^2,
\end{align*}
where $X$ denotes the  ground truth image and  $\widehat{X}$ is the estimated color image.
From $1.0$ to $1.9$, the CPSNR  increases first and then decreases. We can see that the best values are
distributed on the lines with factors from $1.4$ to $1.7$.  A similar behavior was also observed using nlBayes for denoising as well as for other levels of noise (see the supplementary material).

\begin{figure}[ht!]
\begin{center}
\renewcommand{\arraystretch}{0.2} \addtolength{\tabcolsep}{-5.5pt} \vskip3mm {%
\small
\begin{tabular}{ccccH}
\includegraphics[width=0.23\linewidth]{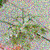}&
\includegraphics[width=0.23\linewidth]{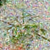}&
\includegraphics[width=0.23\linewidth]{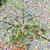}&
\includegraphics[width=0.23\linewidth]{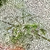}&
\includegraphics[width=0.23\linewidth]{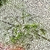}
\\
\includegraphics[width=0.23\linewidth]{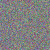}&
\includegraphics[width=0.23\linewidth]{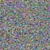}&
\includegraphics[width=0.23\linewidth]{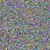}&
\includegraphics[width=0.23\linewidth]{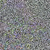}&
\includegraphics[width=0.23\linewidth]{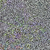}
\\
\includegraphics[width=0.23\linewidth]{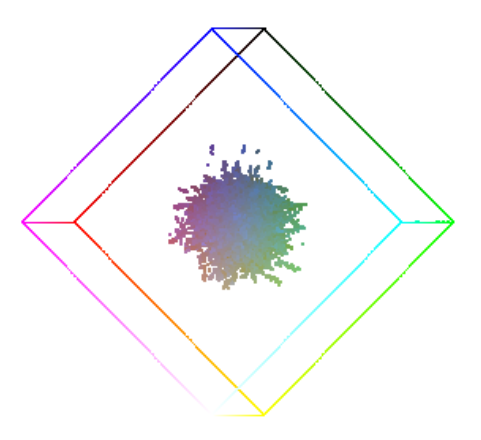}&
\includegraphics[width=0.23\linewidth]{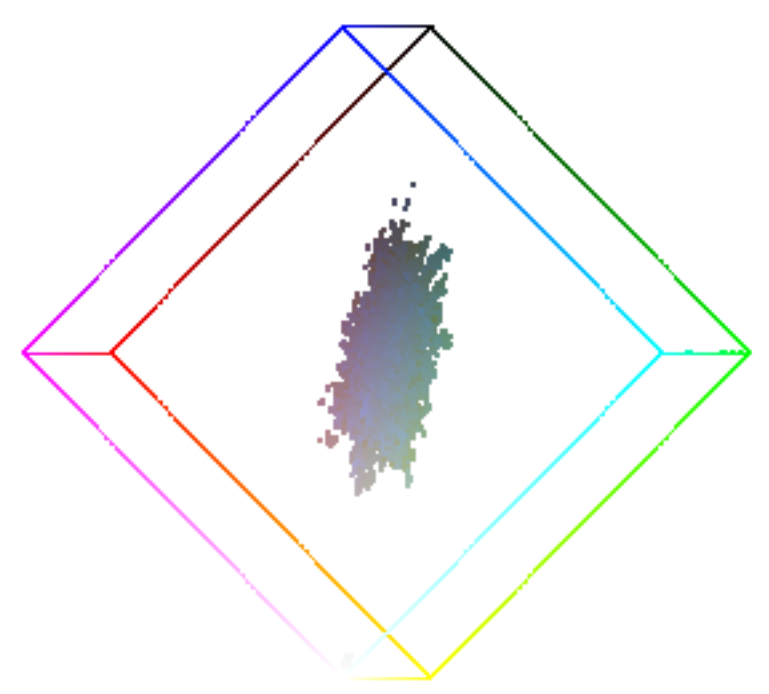}&
\includegraphics[width=0.23\linewidth]{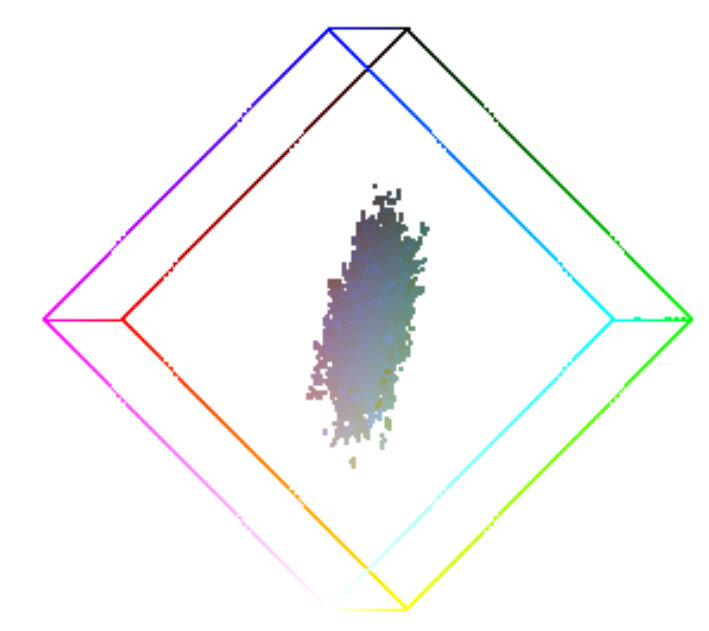}&
\includegraphics[width=0.23\linewidth]{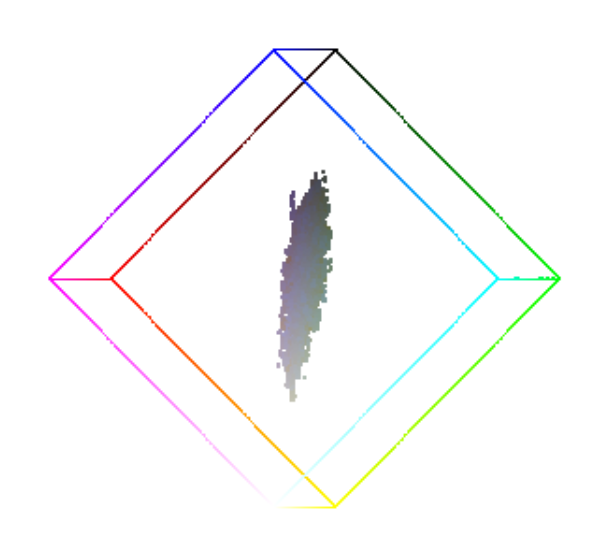}&
\includegraphics[width=0.23\linewidth]{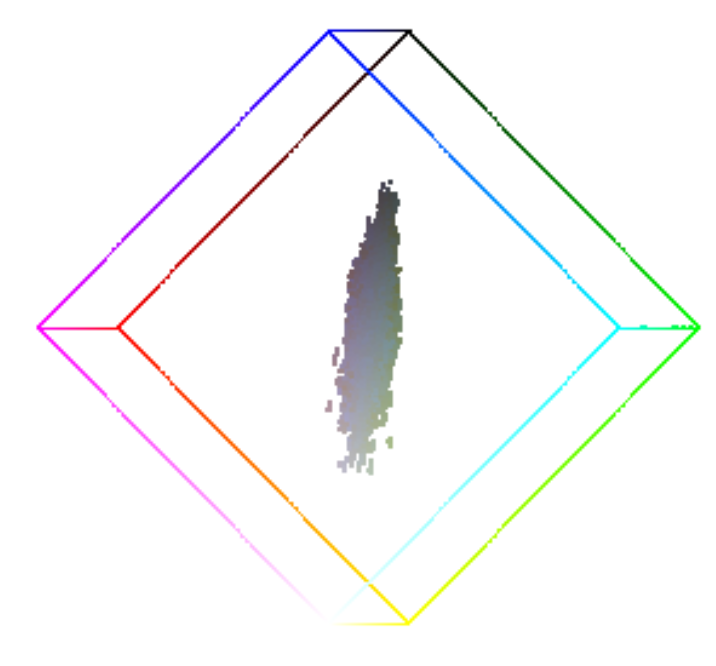}\\
(a) AWG  & (b)  HA   &  (c)  MLRI   &  (d)  RCNN   &  (e)  JCNN
\end{tabular}
} \vskip1mm
\par
\rule{0pt}{-0.2pt}%
\par
\vskip1mm 
\end{center}
\vspace{-.8em}

\caption{{\protect  AWG noise image and demosaicking noise with standard deviation $\sigma = 20$ for respectively HA,  MLRI, RCNN. Last row: the color spaces (in standard (R,G,B) Cartesian coordinates) of each noise, presented in their projection with maximal area.  As expected, the AWG color space is isotropic, while the  color space after demosaicking is elongated in the luminance  direction $Y$  and squeezed in the others.  This amounts to an increased noise standard deviation for $Y$ after demosaicking, and less noise in the chromatic directions.  }}
\label{Fig: noisyimageskind20}
\end{figure}

This does  not  mean that the overall noise standard deviation  has increased
after demosaicking. Let us consider the  noise standard deviation estimated as the
 mean RMSE of the demosaicked images from  the
Imax~\cite{zhang2011color} dataset with different noise levels, given in
Table~\ref{Table rmse}. We observe that for low noise  ($\sigma_0=1$) there is a serious demosaicking error, of about 4, not caused by the noise, but by the
demosaicking itself. However, for   $\sigma_0>10$  we see that the RMSE of the demosaicked image tends to roughly 3/4 of the initial  noise standard deviation.
At  first sight, this $3/4$ factor seems to contradict the observation that denoising with a parameter $1.5\sigma_0$  yields better results. This  leads us to analyzing the structure  of the residual noise.
%
Fig.~\ref{Fig: noisyimageskind20}  shows an image contaminated with AWG noise  with standard deviation $\sigma_0 = 20$ and its resulting demosaicked noise  for respectively HA,  MLRI, RCNN. In the last row of the  figure, one can observe the color spaces (in standard (R,G,B) Cartesian coordinates) of each of these noises, each cloud being presented in its projection with maximal area.  As expected, the AWG color space is isotropic and has an apparent diameter proportional  to  $4\sigma_0\simeq 80$. 
 The color space of the demosaicked noise  is instead elongated in the luminance direction  $Y= \frac{R+G+B}{\sqrt3}$  to about $6\sigma_0\simeq 120$ and squeezed in the others.  This amounts to an increased noise standard deviation for $Y$ after demosaicking, and much less noise in the chromatic directions.

This is confirmed by Table \ref{Table Correlationpixels4level20} that shows variances and covariances of  $(R,G,B)$ and $(Y,U,V)$ respectively for an AWG noise with
$\sigma_0=20$, and  then for the demosaicked noise obtained from it after demosaiciking with RI, MLRI and RCNN.
In Table \ref{Table Correlationpixels4level20} (a) these statistics are computed on a pure  white noise image with $\sigma=20$. Hence the variance of $Y$ is $400$, as the $(R,G,B)\to (Y,U,V)$ transform is  implemented  as an isometry  of $\mathbf{R}^3$.  The variance of  $Y$ is a growing  sequence  for the demosaicked noise obtained by increasingly sophisticated demosaicking: $715$ for RI,  $772$ for MLRI, $972$ for RCNN. \nada{Hence  by these methods the initial Y  noise has its  variance  nearly doubled by RI  and MLRI,  and even  more than doubled by RCNN, from 400 to  972.}{}  In  contrast, the demosaicked noise is reduced in the $U$  and $V$ axes, with its variance passing  from $400$ for AWGN to $168$  and $94$ for RI, and  even  down  to $43$ and  $55$ for RCNN.  Hence, the noise standard deviation on $U$ or $V$ has been divided by a factor between $2$ and $3$.  But Table \ref{Table Correlationpixels4level20} also  shows that the residual noise on $U$ and $V$ is strongly spatially correlated,  it  is therefore a low frequency noise, that will require  stronger \nada{Wiener thresholds}{filtering} than white noise to be removed.  This  table also shows that  the $Y$ component of  the demosaicked noise remains almost  white.  

This leads to a simple conclusion: {since image  denoising algorithms are guided by the $Y$ component~\cite{dabov2007color,lebrun2013nonlocal}, we can denoise with methods designed for white noise, but with a noise parameter  adapted to the increased variance of $Y$.}

To understand why the variance of $Y$ is far larger than the AWG noise it comes from, let us study in Table~\ref{Table Correlation3channels4level20} the correlation between the three  channels $(R, G,  B)$  in the demosaicked noise of RI,  MLRI and RCNN. 
We observe a strong $(R,G,B)$ correlation ranging from 0.6 for RI to 0.89 for RCNN, which is caused by the "tendency to grey" of all demosaicking algorithms. Assuming that the
demosaicked noisy pixel components (denoted 
$\widetilde{\epsilon}_{R},\widetilde{\epsilon}_{G},\widetilde{\epsilon}_{B}$) have a correlation coefficient close to $1$ then we have 
 $$Y=\frac{\widetilde{\epsilon}_{R}+\widetilde{\epsilon}_{G}+\widetilde{\epsilon}_{B}}{\sqrt{3}} \sim \sqrt{3}\, N(0, \sigma_0).$$
This  factor of about $1.7$ corresponds to the case  with maximum correlation. Our empirical observation of an optimal factor near $1.5$ responds to a lower correlation between the colors.

\begin{table}[h!]
\caption{ Variance and covariance of $(R,G,B)$ and $(Y,U,V)$ (each first row) and the  corresponding correlations (each second row) between  pixels $(i,j)$ and $(i+s,j+t)$, $s,t = 0,1,2$ first for AWGN (a) with standard deviation $\sigma = 20$, then for its  demosaicked versions by RI (b),  MLRI (c) and  RCNN (d).  \vspace{-0.25cm}}
\label{Table Correlationpixels4level20}
\begin{center}
\renewcommand{\arraystretch}{1.15} \addtolength{\tabcolsep}{-4.7pt} \vskip3mm
\small

\begin{tabular}{l |rrrrrrrrrr}
      \rowcolor{gray!50}

  & \scriptsize (\emph{i,j}) & \scriptsize  (\emph{i,j}+1) &  \scriptsize (\emph{i,j}+2) &  \scriptsize (\emph{i}+1,\emph{j}) & \scriptsize (\emph{i}+1,\emph{j}+1)&  \scriptsize (\emph{i}+1,\emph{j}+2)  &  \scriptsize (\emph{i}+2,\emph{j})  &  \scriptsize (\emph{i}+2,\emph{j}+1) &  \scriptsize (\emph{i}+2,\emph{j}+2)
   \\\hline\hline

\footnotesize R &\cellcolor[rgb]{0.97 0.67 0.56} 400.6 &\cellcolor[rgb]{0.51 0.65 0.99} 0.6 &\cellcolor[rgb]{0.51 0.64 0.98} 0.4 &\cellcolor[rgb]{0.51 0.65 0.99} 0.7 &\cellcolor[rgb]{0.49 0.63 0.98} 0.1 &\cellcolor[rgb]{0.51 0.65 0.99} 0.7 &\cellcolor[rgb]{0.50 0.64 0.98} 0.3 &\cellcolor[rgb]{0.50 0.64 0.98} 0.2 &\cellcolor[rgb]{0.52 0.65 0.99} 0.8\\
\footnotesize G &\cellcolor[rgb]{0.97 0.67 0.56} 401.7 &\cellcolor[rgb]{0.51 0.64 0.98} 0.5 &\cellcolor[rgb]{0.52 0.65 0.99} 1.1 &\cellcolor[rgb]{0.49 0.63 0.98} 0.1 &\cellcolor[rgb]{0.50 0.64 0.98} 0.3 &\cellcolor[rgb]{0.52 0.65 0.99} 0.9 &\cellcolor[rgb]{0.52 0.65 0.99} 1.0 &\cellcolor[rgb]{0.51 0.65 0.99} 0.6 &\cellcolor[rgb]{0.51 0.64 0.98} 0.4\\
\footnotesize B &\cellcolor[rgb]{0.97 0.67 0.56} 400.2 &\cellcolor[rgb]{0.52 0.66 0.99} 1.2 &\cellcolor[rgb]{0.49 0.63 0.98} 0.1 &\cellcolor[rgb]{0.51 0.64 0.98} 0.5 &\cellcolor[rgb]{0.51 0.65 0.99} 0.6 &\cellcolor[rgb]{0.49 0.63 0.98} 0.0 &\cellcolor[rgb]{0.53 0.66 0.99} 1.9 &\cellcolor[rgb]{0.50 0.64 0.98} 0.3 &\cellcolor[rgb]{0.53 0.66 0.99} 1.9\\
\hline\hline
\footnotesize Y &\cellcolor[rgb]{0.97 0.67 0.56} 399.6 &\cellcolor[rgb]{0.52 0.65 0.99} 1.1 &\cellcolor[rgb]{0.49 0.63 0.98} 0.1 &\cellcolor[rgb]{0.50 0.64 0.98} 0.3 &\cellcolor[rgb]{0.49 0.63 0.98} 0.1 &\cellcolor[rgb]{0.52 0.65 0.99} 0.9 &\cellcolor[rgb]{0.50 0.64 0.98} 0.2 &\cellcolor[rgb]{0.51 0.64 0.98} 0.5 &\cellcolor[rgb]{0.52 0.66 0.99} 1.2\\
\footnotesize U &\cellcolor[rgb]{0.97 0.67 0.56} 401.5 &\cellcolor[rgb]{0.49 0.63 0.98} 0.1 &\cellcolor[rgb]{0.52 0.65 0.99} 0.8 &\cellcolor[rgb]{0.51 0.65 0.99} 0.6 &\cellcolor[rgb]{0.50 0.64 0.98} 0.3 &\cellcolor[rgb]{0.50 0.64 0.98} 0.3 &\cellcolor[rgb]{0.52 0.65 0.99} 0.9 &\cellcolor[rgb]{0.51 0.64 0.98} 0.5 &\cellcolor[rgb]{0.52 0.66 0.99} 1.3\\
\footnotesize V &\cellcolor[rgb]{0.97 0.67 0.56} 401.4 &\cellcolor[rgb]{0.49 0.63 0.98} 0.2 &\cellcolor[rgb]{0.53 0.66 0.99} 1.8 &\cellcolor[rgb]{0.52 0.65 0.99} 0.9 &\cellcolor[rgb]{0.50 0.64 0.98} 0.2 &\cellcolor[rgb]{0.52 0.65 0.99} 1.0 &\cellcolor[rgb]{0.51 0.65 0.99} 0.6 &\cellcolor[rgb]{0.50 0.64 0.98} 0.2 &\cellcolor[rgb]{0.50 0.64 0.98} 0.2\\
\hline
  \multicolumn{10}{c}{ (a)  AWG noise   } \\
        \rowcolor{gray!50}
  & \scriptsize (\emph{i,j}) & \scriptsize  (\emph{i,j}+1) &  \scriptsize (\emph{i,j}+2) &  \scriptsize (\emph{i}+1,\emph{j}) & \scriptsize (\emph{i}+1,\emph{j}+1)&  \scriptsize (\emph{i}+1,\emph{j}+2)  &  \scriptsize (\emph{i}+2,\emph{j})  &  \scriptsize (\emph{i}+2,\emph{j}+1) &  \scriptsize (\emph{i}+2,\emph{j}+2)
   \\\hline\hline

\footnotesize R &\cellcolor[rgb]{0.97 0.73 0.62} 336.4 &\cellcolor[rgb]{0.85 0.86 0.89} 126.8 &\cellcolor[rgb]{0.64 0.76 1.00} 19.4 &\cellcolor[rgb]{0.85 0.86 0.88} 129.9 &\cellcolor[rgb]{0.73 0.82 0.97} 52.9 &\cellcolor[rgb]{0.64 0.76 1.00} 21.6 &\cellcolor[rgb]{0.64 0.76 1.00} 20.7 &\cellcolor[rgb]{0.65 0.76 1.00} 22.4 &\cellcolor[rgb]{0.63 0.75 1.00} 18.7\\
\footnotesize G &\cellcolor[rgb]{0.96 0.76 0.67} 295.5 &\cellcolor[rgb]{0.80 0.85 0.93} 92.5 &\cellcolor[rgb]{0.51 0.64 0.98} 0.5 &\cellcolor[rgb]{0.80 0.85 0.93} 95.6 &\cellcolor[rgb]{0.64 0.76 1.00} 20.6 &\cellcolor[rgb]{0.53 0.66 0.99} 1.8 &\cellcolor[rgb]{0.51 0.65 0.99} 0.7 &\cellcolor[rgb]{0.52 0.66 0.99} 1.5 &\cellcolor[rgb]{0.55 0.69 1.00} 4.3\\
\footnotesize B &\cellcolor[rgb]{0.97 0.72 0.61} 350.5 &\cellcolor[rgb]{0.85 0.86 0.89} 125.9 &\cellcolor[rgb]{0.63 0.75 1.00} 18.1 &\cellcolor[rgb]{0.85 0.86 0.88} 130.4 &\cellcolor[rgb]{0.72 0.81 0.98} 50.7 &\cellcolor[rgb]{0.64 0.76 1.00} 20.8 &\cellcolor[rgb]{0.64 0.76 1.00} 20.0 &\cellcolor[rgb]{0.64 0.76 1.00} 20.9 &\cellcolor[rgb]{0.62 0.75 1.00} 17.5\\
\hline\hline
\footnotesize Y &\cellcolor[rgb]{0.87 0.39 0.31} 715.6 &\cellcolor[rgb]{0.89 0.85 0.83} 170.9 &\cellcolor[rgb]{0.68 0.79 0.99} 32.3 &\cellcolor[rgb]{0.90 0.85 0.82} 178.6 &\cellcolor[rgb]{0.54 0.67 0.99} 2.6 &\cellcolor[rgb]{0.57 0.70 1.00} 5.4 &\cellcolor[rgb]{0.68 0.79 0.99} 34.0 &\cellcolor[rgb]{0.58 0.71 1.00} 7.1 &\cellcolor[rgb]{0.64 0.76 1.00} 20.5\\
\footnotesize U &\cellcolor[rgb]{0.89 0.85 0.83} 168.4 &\cellcolor[rgb]{0.82 0.86 0.91} 108.3 &\cellcolor[rgb]{0.70 0.80 0.98} 41.3 &\cellcolor[rgb]{0.83 0.86 0.91} 110.1 &\cellcolor[rgb]{0.77 0.84 0.95} 73.4 &\cellcolor[rgb]{0.67 0.78 0.99} 28.2 &\cellcolor[rgb]{0.71 0.81 0.98} 44.1 &\cellcolor[rgb]{0.67 0.78 0.99} 29.4 &\cellcolor[rgb]{0.59 0.72 1.00} 9.7\\
\footnotesize V &\cellcolor[rgb]{0.81 0.85 0.92} 98.3 &\cellcolor[rgb]{0.76 0.83 0.96} 66.0 &\cellcolor[rgb]{0.67 0.78 0.99} 27.9 &\cellcolor[rgb]{0.76 0.83 0.96} 67.3 &\cellcolor[rgb]{0.72 0.81 0.98} 48.1 &\cellcolor[rgb]{0.64 0.76 1.00} 21.4 &\cellcolor[rgb]{0.67 0.78 0.99} 29.9 &\cellcolor[rgb]{0.65 0.76 1.00} 22.4 &\cellcolor[rgb]{0.59 0.72 1.00} 10.4\\
\hline
  \multicolumn{10}{c}{ (b)  RI  } \\
      \rowcolor{gray!50}
  & \scriptsize (\emph{i,j}) & \scriptsize  (\emph{i,j}+1) &  \scriptsize (\emph{i,j}+2) &  \scriptsize (\emph{i}+1,\emph{j}) & \scriptsize (\emph{i}+1,\emph{j}+1)&  \scriptsize (\emph{i}+1,\emph{j}+2)  &  \scriptsize (\emph{i}+2,\emph{j})  &  \scriptsize (\emph{i}+2,\emph{j}+1) &  \scriptsize (\emph{i}+2,\emph{j}+2)
   \\\hline\hline

\footnotesize R &\cellcolor[rgb]{0.97 0.71 0.60} 361.4 &\cellcolor[rgb]{0.85 0.86 0.89} 128.4 &\cellcolor[rgb]{0.64 0.76 1.00} 18.9 &\cellcolor[rgb]{0.85 0.86 0.88} 130.5 &\cellcolor[rgb]{0.71 0.81 0.98} 46.4 &\cellcolor[rgb]{0.64 0.76 1.00} 20.6 &\cellcolor[rgb]{0.64 0.76 1.00} 21.6 &\cellcolor[rgb]{0.64 0.76 1.00} 21.5 &\cellcolor[rgb]{0.64 0.76 1.00} 19.8\\
\footnotesize G &\cellcolor[rgb]{0.96 0.76 0.67} 298.9 &\cellcolor[rgb]{0.80 0.85 0.93} 93.0 &\cellcolor[rgb]{0.51 0.64 0.98} 0.5 &\cellcolor[rgb]{0.80 0.85 0.93} 95.1 &\cellcolor[rgb]{0.64 0.76 1.00} 19.1 &\cellcolor[rgb]{0.52 0.65 0.99} 0.9 &\cellcolor[rgb]{0.52 0.65 0.99} 1.0 &\cellcolor[rgb]{0.51 0.64 0.98} 0.5 &\cellcolor[rgb]{0.55 0.69 0.99} 3.8\\
\footnotesize B &\cellcolor[rgb]{0.97 0.70 0.59} 370.9 &\cellcolor[rgb]{0.85 0.86 0.89} 127.8 &\cellcolor[rgb]{0.64 0.76 1.00} 19.3 &\cellcolor[rgb]{0.85 0.86 0.88} 130.4 &\cellcolor[rgb]{0.71 0.81 0.98} 46.0 &\cellcolor[rgb]{0.64 0.76 1.00} 20.6 &\cellcolor[rgb]{0.64 0.76 1.00} 21.2 &\cellcolor[rgb]{0.64 0.76 1.00} 20.3 &\cellcolor[rgb]{0.64 0.76 1.00} 19.0\\
\hline\hline
\footnotesize Y &\cellcolor[rgb]{0.84 0.33 0.27} 772.2 &\cellcolor[rgb]{0.90 0.85 0.82} 177.7 &\cellcolor[rgb]{0.68 0.79 0.99} 33.0 &\cellcolor[rgb]{0.90 0.84 0.81} 181.3 &\cellcolor[rgb]{0.59 0.72 1.00} 9.6 &\cellcolor[rgb]{0.59 0.72 1.00} 9.2 &\cellcolor[rgb]{0.68 0.79 0.99} 32.6 &\cellcolor[rgb]{0.60 0.73 1.00} 10.9 &\cellcolor[rgb]{0.64 0.76 1.00} 21.4\\
\footnotesize U &\cellcolor[rgb]{0.89 0.85 0.83} 164.8 &\cellcolor[rgb]{0.82 0.86 0.91} 107.1 &\cellcolor[rgb]{0.71 0.81 0.98} 43.7 &\cellcolor[rgb]{0.82 0.86 0.91} 108.8 &\cellcolor[rgb]{0.77 0.84 0.95} 72.8 &\cellcolor[rgb]{0.67 0.78 0.99} 29.3 &\cellcolor[rgb]{0.71 0.81 0.98} 46.1 &\cellcolor[rgb]{0.67 0.78 0.99} 30.2 &\cellcolor[rgb]{0.59 0.72 1.00} 10.1\\
\footnotesize V &\cellcolor[rgb]{0.80 0.85 0.93} 94.3 &\cellcolor[rgb]{0.75 0.83 0.96} 64.4 &\cellcolor[rgb]{0.67 0.78 0.99} 28.1 &\cellcolor[rgb]{0.76 0.83 0.96} 65.8 &\cellcolor[rgb]{0.72 0.81 0.98} 48.2 &\cellcolor[rgb]{0.65 0.76 1.00} 21.9 &\cellcolor[rgb]{0.67 0.78 0.99} 30.3 &\cellcolor[rgb]{0.65 0.76 1.00} 23.1 &\cellcolor[rgb]{0.60 0.73 1.00} 11.1\\
\hline
  \multicolumn{10}{c}{ (c)  MLRI   } \\
       \rowcolor{gray!50}
  & \scriptsize (\emph{i,j}) & \scriptsize  (\emph{i,j}+1) &  \scriptsize (\emph{i,j}+2) &  \scriptsize (\emph{i}+1,\emph{j}) & \scriptsize (\emph{i}+1,\emph{j}+1)&  \scriptsize (\emph{i}+1,\emph{j}+2)  &  \scriptsize (\emph{i}+2,\emph{j})  &  \scriptsize (\emph{i}+2,\emph{j}+1) &  \scriptsize (\emph{i}+2,\emph{j}+2)
   \\\hline\hline

\footnotesize R &\cellcolor[rgb]{0.97 0.71 0.60} 359.9 &\cellcolor[rgb]{0.72 0.81 0.98} 47.8 &\cellcolor[rgb]{0.56 0.69 1.00} 5.0 &\cellcolor[rgb]{0.73 0.82 0.97} 51.9 &\cellcolor[rgb]{0.65 0.76 1.00} 21.8 &\cellcolor[rgb]{0.63 0.75 1.00} 17.8 &\cellcolor[rgb]{0.56 0.69 1.00} 5.1 &\cellcolor[rgb]{0.64 0.76 1.00} 19.4 &\cellcolor[rgb]{0.59 0.72 1.00} 9.2\\
\footnotesize G &\cellcolor[rgb]{0.97 0.72 0.61} 354.8 &\cellcolor[rgb]{0.68 0.79 0.99} 32.6 &\cellcolor[rgb]{0.55 0.69 1.00} 4.4 &\cellcolor[rgb]{0.69 0.79 0.99} 36.3 &\cellcolor[rgb]{0.57 0.70 1.00} 5.8 &\cellcolor[rgb]{0.58 0.71 1.00} 8.4 &\cellcolor[rgb]{0.57 0.70 1.00} 6.4 &\cellcolor[rgb]{0.59 0.72 1.00} 8.8 &\cellcolor[rgb]{0.51 0.64 0.98} 0.6\\
\footnotesize B &\cellcolor[rgb]{0.97 0.72 0.61} 356.0 &\cellcolor[rgb]{0.72 0.81 0.98} 49.6 &\cellcolor[rgb]{0.57 0.70 1.00} 6.3 &\cellcolor[rgb]{0.73 0.82 0.97} 53.7 &\cellcolor[rgb]{0.65 0.77 1.00} 23.6 &\cellcolor[rgb]{0.63 0.75 1.00} 18.8 &\cellcolor[rgb]{0.58 0.71 1.00} 7.3 &\cellcolor[rgb]{0.64 0.76 1.00} 19.4 &\cellcolor[rgb]{0.59 0.72 1.00} 9.2\\
\hline\hline
\footnotesize Y &\cellcolor[rgb]{0.74 0.10 0.17} 972.3 &\cellcolor[rgb]{0.76 0.84 0.96} 69.0 &\cellcolor[rgb]{0.64 0.76 1.00} 20.8 &\cellcolor[rgb]{0.78 0.84 0.95} 76.4 &\cellcolor[rgb]{0.55 0.69 0.99} 3.6 &\cellcolor[rgb]{0.63 0.75 1.00} 18.6 &\cellcolor[rgb]{0.67 0.78 0.99} 28.9 &\cellcolor[rgb]{0.62 0.75 1.00} 17.3 &\cellcolor[rgb]{0.53 0.67 0.99} 2.2\\
\footnotesize U &\cellcolor[rgb]{0.73 0.82 0.97} 55.1 &\cellcolor[rgb]{0.68 0.79 0.99} 33.8 &\cellcolor[rgb]{0.62 0.74 1.00} 15.3 &\cellcolor[rgb]{0.69 0.79 0.99} 36.0 &\cellcolor[rgb]{0.66 0.77 0.99} 26.1 &\cellcolor[rgb]{0.61 0.74 1.00} 14.6 &\cellcolor[rgb]{0.64 0.76 1.00} 19.0 &\cellcolor[rgb]{0.62 0.75 1.00} 16.6 &\cellcolor[rgb]{0.60 0.73 1.00} 11.8\\
\footnotesize V &\cellcolor[rgb]{0.71 0.81 0.98} 43.3 &\cellcolor[rgb]{0.66 0.78 0.99} 27.3 &\cellcolor[rgb]{0.60 0.73 1.00} 12.3 &\cellcolor[rgb]{0.67 0.78 0.99} 29.4 &\cellcolor[rgb]{0.64 0.76 1.00} 21.5 &\cellcolor[rgb]{0.60 0.73 1.00} 11.7 &\cellcolor[rgb]{0.62 0.74 1.00} 16.0 &\cellcolor[rgb]{0.61 0.74 1.00} 13.7 &\cellcolor[rgb]{0.59 0.72 1.00} 9.4\\
\hline
  \multicolumn{10}{c}{ (d)  RCNN   }

\end{tabular}
\end{center}
\end{table}

\begin{table}[t]
\caption{ Covariances (each first row) and \emph{correlations} (each second row) of the three color  channels (R, G, B) of the  demosaicked noise, when the initial CFA white noise satisfies $\sigma_0 = 20$   }
\label{Table Correlation3channels4level20}

\vspace{-.8em}

\begin{center}
\renewcommand{\arraystretch}{1.25} \addtolength{\tabcolsep}{-2pt} \vskip3mm
\small

\begin{minipage}[b]{0.5\linewidth}
   \centering
   
\begin{tabular}{c|ccc}
  & R & G & B \\  
  \hline
\multirow{2}*{R}& 336.44& 206.29& 175.01 \\
& \em 1.0000& \em 0.6542& \em 0.5097\\
\hline 
\multirow{2}*{G}& 206.29& 295.54& 200.96\\
& \em 0.6542& \em 1.0000& \em 0.6244\\
\hline
\multirow{2}*{B}& 175.01& 200.96& 350.46\\
& \em 0.5097& \em 0.6244& \em 1.0000\\
\hline
\end{tabular}
\medskip

(a)  RI 
\smallskip
  \end{minipage}%
\begin{minipage}[b]{0.5\linewidth}
   \centering
   
\begin{tabular}{c|ccc}
  & R & G & B \\  
  \hline
\multirow{2}*{R}& 361.42& 224.39& 201.41\\
& \em 1.0000& \em 0.6826& \em 0.5501\\
\hline 
\multirow{2}*{G}& 224.39& 298.94& 216.86\\
& \em 0.6826& \em 1.0000& \em 0.6512\\
\hline
\multirow{2}*{B}& 201.41& 216.86& 370.92\\
& \em 0.5501&\em  0.6512& \em 1.0000\\
\hline
\end{tabular}
\medskip

(b) MLRI
\smallskip

  \end{minipage}
\begin{minipage}[b]{0.5\linewidth}
   \centering
   
\begin{tabular}{c|ccc}
  & R & G & B \\  
  \hline
\multirow{2}*{R}& 359.90& 320.44& 302.85\\
& \em 1.0000& \em 0.8967& \em 0.8461\\
\hline 
\multirow{2}*{G}& 320.44& 354.83& 299.85\\
& \em 0.8967& \em 1.0000& \em 0.8437\\
\hline
\multirow{2}*{B}& 302.85& 299.85& 355.99\\
& \em 0.8461& \em 0.8437& \em 1.0000\\
\hline
\end{tabular}
\medskip

(c) RCNN
\smallskip

  \end{minipage}%
\begin{minipage}[b]{0.5\linewidth}
   \centering
   
\begin{tabular}{c|ccc}
  & R & G & B \\  
  \hline
\multirow{2}*{R}& 334.84& 297.31& 275.28\\
& \em 1.0000& \em 0.8675& \em 0.8181\\
\hline 
\multirow{2}*{G}& 297.31& 350.81& 270.32\\
& \em 0.8675& \em 1.0000& \em 0.7848\\
\hline
\multirow{2}*{B}& 275.28& 270.32& 338.17\\
& \em 0.8181& \em 0.7848& \em 1.0000\\
\hline
\end{tabular}
\medskip

(d) JCNN
\smallskip

  \end{minipage}

\end{center}

\end{table}

\begin{table}[!htbp]
\caption{Comparison in CPSNR(dB)  of average restoration performance between $\A\&\B$ and $\B\&\A$ for a fixed level of noise $\sigma_0=20$.  
We test two denoisers $\A$ namely CBM3D, and nlBayes, and  $1.5\A$ means that if noise level is $\sigma_0$,  the noise level parameter for the  denoising method $\A$ is $\sigma=1.5\sigma_0$. 
Both denoisers can be adapted to handle mosaics  in the $\A\&\B$ schemes (see in the text).
The best result of each column  is marked with a \fbox{box}.  The best result of each line is in \textcolor{red}{\bf red} and the second best one is in \textcolor{green}{\bf green}. 
}
\vspace{-0.25cm}
\label{Table coI}
\begin{center}
\renewcommand{\arraystretch}{1.25} \addtolength{\tabcolsep}{-5.8pt} \vskip3mm
\small
\begin{tabular}{l| l| c H c c H c H c H H}
      \rowcolor{gray!50}
\hline
$\A$ &Algorithm&HA&GBTF&RI&MLRI&MLRI&ARI&LSSC&RCNN&JCNN& Ave
 \\\hline

\multirow{4}{*}{
\rotatebox[origin=l]{90}{CBM3D}
}
&  $\A\&\B$&28.11& 0& 28.45& 27.97& 0& {\color{red} \bf 28.69 }& \fbox{0}& { 27.27}& { \color{green} \bf 27.86}&28.06\\
&  $\B\&\A$&28.15& 27.58& { \color{green}{\bf 28.46}}& 27.95& 27.83& { \color{red} {\bf 28.70}}& 27.19& 27.28& 27.89&28.07\\
&  $\B\&1.5 \A$&\fbox{29.24}& 29.01& \fbox{29.32}& \fbox{29.22}& 29.15& \fbox{{ {29.36}}}& \fbox{29.00}& \fbox{\color{green} \bf  29.41}&  \fbox{{ \color{red} {\bf 29.59}}}&\fbox{29.36}\\
\hline
\multirow{4}{*}{
\rotatebox[origin=l]{90}{nlBayes}
}& $\A\&\B$&28.17& 28.00& 28.17& 28.17& 28.18& 28.18& 28.18& { \color{red} {\bf 28.28}}& { \color{green}{\bf 28.19}}&28.19\\
& $\B\&\A$&28.67& 28.15& { \color{green}{\bf 28.99}}& 28.57& 28.45& { \color{red} {\bf 29.21}}& 27.82& 28.02& 28.51&28.66\\
& $\B\&1.5\A$&\fbox{29.29}& 28.99& \fbox{29.26}& \fbox{29.22}& 29.23& \fbox{29.31}& \fbox{29.04}& \fbox{ \color{green}{\bf 29.36}}& \fbox{ \color{red} {\bf 29.45}}&\fbox{29.32}\\
\hline

\end{tabular}
\end{center}
\end{table}

\section{Experimental evaluation}
\label{sec:experiment}

To evaluate the proposed framework for denoising and demosaicking, we shall use two classic noise free  color image datasets: Kodak  and Imax.
The Imax dataset~\cite{zhang2011color} consists  of 18  images of $500\times 500$ pixels, cropped from   high-resolution images of size: $2310\times 1814$.
The Kodak  dataset  consists of 25  images of  $768 \times 512$ pixels released by the Kodak Corporation for unrestricted research usage\footnote{Image source: \url{http://r0k.us/graphics/kodak}}.
\gf{}{We also evaluated it on a set of 14 real raw images  from the SIDD dataset~\cite{Abdelhamed2018}, which comes with ground truth acquisitions.}

\paragraph{Evaluation of \A\&\B~and \B\&\A~strategies.}
We performed simulations with \nada{three different schemes:  $\A\&\B$, $\B\&\A$ and  $\B\&A\&\B$. Here $\B\&A\&\B$ is a three-step scheme which adds a demosaicking step after the $\B\&\A$ scheme. Specifically, let $(\widetilde{R},\widetilde{G},\widetilde{B})$ be the output of the $\B\&\A$ scheme, then this  result is mosaicked again to get a denoised CFA image $(M_{R}.*\widetilde{R},M_{G}.*\widetilde{G},M_{B}.*\widetilde{B})$. The final  full-color image is obtained by using the same demosaicking algorithm as in the first step.}{the   schemes:  $\A\&\B$ and  $\B\&\A$.}
The considered demosaicking methods  range from classic to very modern:   HA\cite{hamilton1997adaptive}, RI\cite{kiku2013residual}, MLRI\cite{kiku2014minimized},  ARI \cite{monno2017adaptive}, and
RCNN\cite{tan2017color}.
For the denoising stage two classic hand-crafted patch-based denoising algorithms were considered:
 CBM3D~\cite{dabov2007color} and nlBayes~\cite{lebrun2013nonlocal}. As commented in the introduction, both  methods can be adapted to handle mosaics (in the $\A\&\B$ setting). In the case of CBM3D this amounts to  applying the method by Danielyan \emph{et al.}~\cite{danielyan2009cross}, while  for nlBayes this is done by denoising the  4-channel image associated to the mosaic.


   The denoising and demosaicking schemes with the above mentioned demosaicking algorithms and denoising methods were applied to the mosaic images of the  Imax image dataset corrupted by additive white Gaussian noise with standard deviations $\sigma_0 = 1,3,5,10,15,20,25,30,35,40,45,50,55,60$.

    Due to space constraints, in Table~\ref{Table coI} we only report  the 
    results corresponding to one noise level $\sigma_0=20$. Results corresponding to other noise levels are in the supplementary material. 
     From Table~\ref{Table coI}, we can see that $\B\&\A$ with parameter $\sigma = \sigma_0$ is not better than $\A\&\B$, but $\B\&1.5\A$ (which denotes denoising $\A$ with parameter $\sigma = 1.5\sigma_0$) beats  clearly $\A\&\B$.
     {This might explain why} many researchers think that the scheme $\A\&\B$  {was superior to} the scheme $\B\&\A$.

    \nada{ 
    More in detail, the CPSNRs for the $\B\&1.5\A$ schemes surpass the $\A\&\B$ schemes by more than 0.4dB for $\sigma_0 = 5$, and by 0.6dB  for $\sigma_0 = 10$.
     Starting at $\sigma_0 \geq 20$, CBM3D and nlBayes exhibit different results. The $\B\&1.5\A$ scheme with nlBayes outperforms the $\A\&\B$ scheme with nlBayes by  1.09dB on average, while the $\B\&1.5\A$ scheme with CBM3D only wins 0.38dB for $\sigma_0 = 20$. Furthermore, the $\B\&1.5\A$ scheme with CBM3D loses to $\A\&\B$ by  0.27dB on average, but the best  $\B\&\A$ gets CPSNR$=25.95$ and is higher than  $\A\&\B$ with CPSNR$25.82$ when $\sigma_0=40$. The $\B\&1.5\A$ scheme with nlBayes always outperforms $\A\&\B$.  
}


In addition to the good CPSNR results, one important advantage
of the $\B\&\A$ schemes is the high
visual quality of the final restored images.
Fig.~\ref{Fig: bookp1} demonstrates the differences between the various solutions (based on BM3D) obtained on the test image number 3 of the Imax dataset with $\sigma_0=20$.
To save space, only crops of the  full-color results and corresponding differences with the ground truth are shown here.

The  $\A\& \B$ scheme shown in Fig.~\ref{Fig: bookp1} (b1) and (b2) uses BM3D-CFA~\cite{danielyan2009cross} for denoising;  we can observe some minor checkerboard artifacts.
%
From Fig.~\ref{Fig: bookp1} (c1)  and (c2), we can deduce that there is no checkerboard effect but that much noise remains in the restored image by
$\B\&\A$ schemes with parameter $1.0\sigma_0$. 
The result of  $\B\&1.5 \A$ (Fig.~\ref{Fig: bookp1} (d1) and (d2)) are smooth without checkerboard
effects.
%
Fig.~\ref{Fig: bookp1} (e1) and  (e2) correspond to the outputs of the CNN joint denoising and  demosaicking method JCNN~\cite{gharbi2016deep}.

One can observe 
 thin structures 
in the upper left corner of Fig~\ref{Fig: bookp1} (a1), but they disappear in the restored image by $\A\&\B$. The proposed  $\B\&1.5 \A$ scheme restores them. 
The second column of Fig~\ref{Fig: bookp1}  illustrates a similar situation in which thin details are recovered by  $\B\&\A$ and $\B\&1.5 \A$ but  not in the others. 

In short,  it appears  that the $\B\&\A$   scheme with an appropriate parameter (namely $\B\&1.5 \A$)
outperforms the competition in terms of visual quality. This is due to the fact that
it efficiently uses spatial and spectral image
characteristics to remove noise, preserve edges and fine
details. Indeed, contrary to the $\A\&\B$ schemes, $\B\&1.5 \A$ does  not reduce the resolution of the noisy image. Using an $\A\&\B$ scheme ends up over-smoothing the result.
It comes to no surprise that JCNN  performs slightly better than    the other methods; however, it is  much more computationally demanding and only works for $\sigma\leq 20$.

In a systematic comparison between the schemes involving CBM3D and nlBayes, schemes with CBM3D proved to perform slightly better. Furthermore, the $\B\&\A$ schemes with CBM3D are about four times faster than nlBayes. Hence, the following experiments are more focused on  CBM3D.

\nada{ We performed a systematic comparison between the schemes RCNN demosaicking \cite{tan2017color} + CBM3D \cite{dabov2006image} and RCNN demosaicking \cite{tan2017color} + nlBayes \cite{lebrun2013nonlocal}   with $1.5\sigma$. The schemes with CBM3D proved to perform slightly better than the schemes with nlBayes. Hence, from now on, we only show the results of the scheme with CBM3D (see Tables \ref{Table bm3d3cImaxYRGB} and \ref{Table bm3d3cKodakYRGB}). Furthermore, the running times of the $\B\&\A$ schemes for CBM3D $\times$ \{MLRI or JCNN or RCNN\} is given in Table \ref{Table times}. We can see that  $\B\&\A$ schemes with CBM3D are very fast. Tables including the results obtained with nlBayes can be found in the supplementary material.}{}

\begin{figure}[t]
\footnotesize
    \centering
  \addtolength{\tabcolsep}{-5pt} 

    \begin{tabular}{cc}
    \includegraphics[width=.48\linewidth,trim={0 0 0 0},clip]{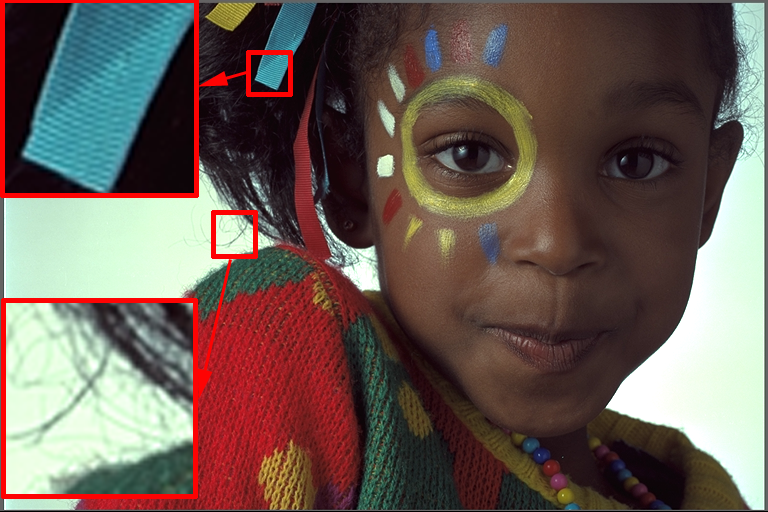}
    & 
\overimg[width=.48\linewidth,trim={0 0 0 0},clip]{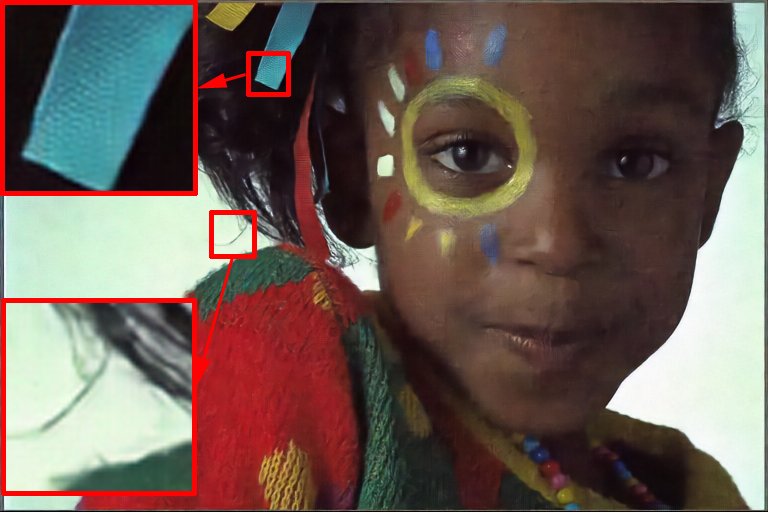}{30.84dB}\\ 

     Ground Truth & 
     JCNN~\cite{gharbi2016deep}  \\
     
\overimg[width=.48\linewidth,trim={0 0 0 0},clip]{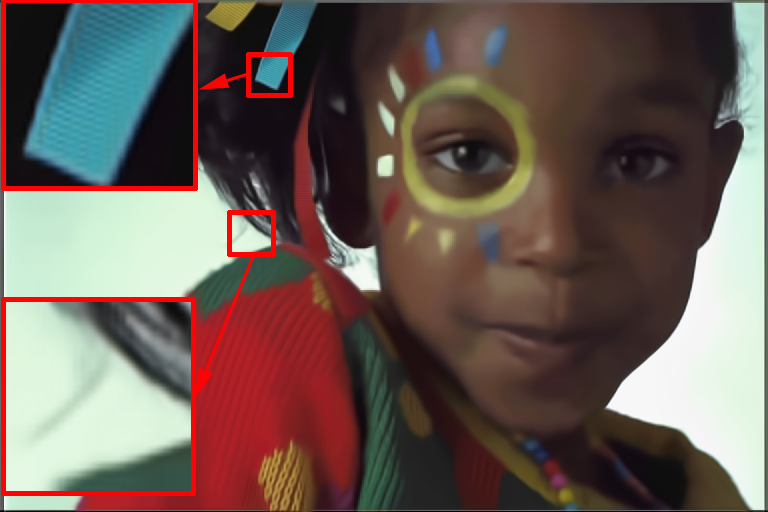}{29.46dB}
& 
\overimg[width=.48\linewidth,trim={0 0 0 0},clip]{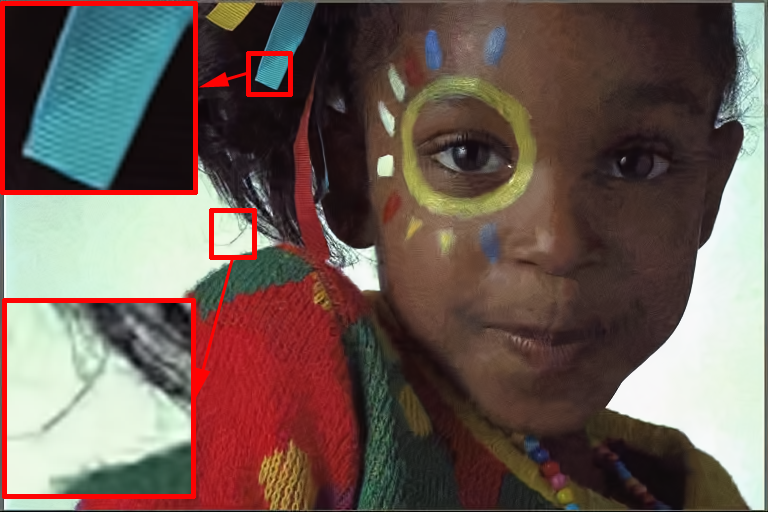}{30.97dB} 

\\ 

     BM3D+RCNN ($\A\&\B$) & RCNN+BM3D ($\B\&1.5\A$)
     \\
\overimg[width=.48\linewidth,trim={0 0 0 0},clip]{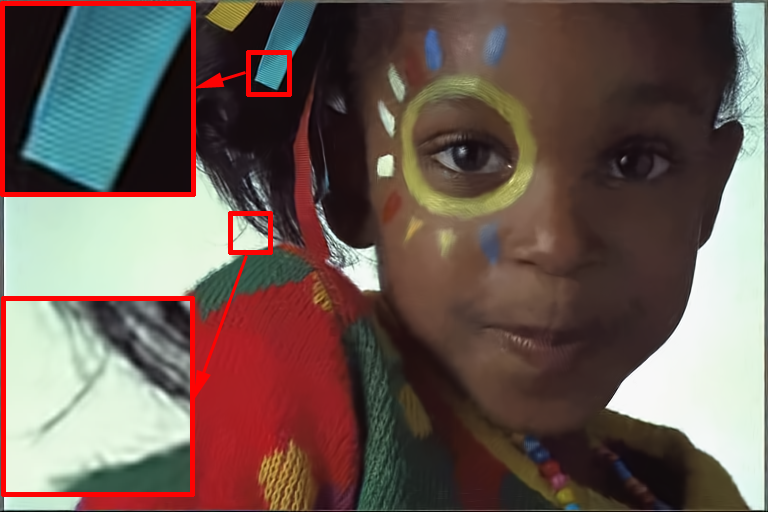}{30.77dB}
&
\overimg[width=.48\linewidth,trim={0 0 0 0},clip]{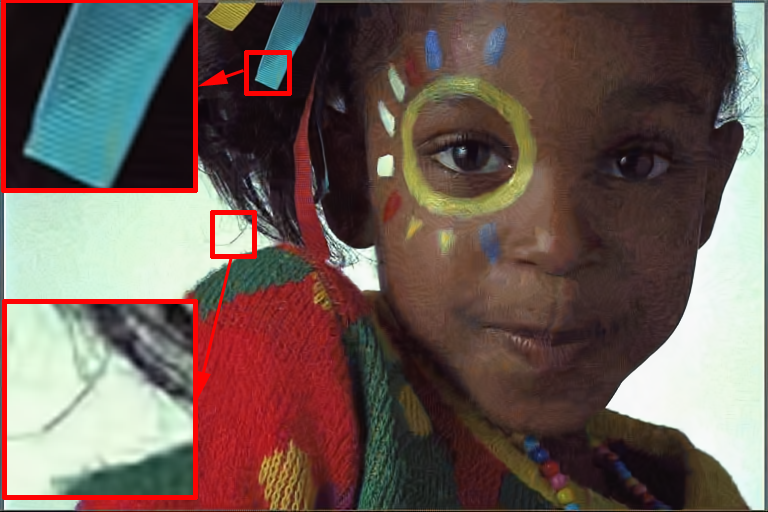}{30.84dB}\\ 

      RCNN+nlBayes  ($\B\&1.5\A$) &     MLRI+BM3D ($\B\&1.5\A$) 
    \end{tabular}

    \label{fig:compare_kodak}
\caption{Demosaicking and denoising results on an image from the Kodak dataset  with $\sigma=20$. We compare an $\A\&\B$ scheme BM3D+RCNN~\cite{danielyan2009cross}, with three $\B\&1.5\A$ RCNN+CBM3D, RCNN+nlBayes and     MLRI+BM3D. As a reference we also include the result of a joint CNN method JCNN~\cite{gharbi2016deep}. But  its results are only available for noise with $\sigma\le20$ because  the network is not trained beyond that level. }
\end{figure}

\paragraph{Comparison with methods from the literature.}

To complete  this  comparison we went  back to all  $\A\&\B$ schemes proposed in the  literature, and performed a systematic comparison for the  two classic Kodak and Imax datasets. These databases are always used in demosaicking evaluations, because they illustrate different challenges of the  demosaicking problem, Imax being difficult by its color contrast, and Kodak challenging for  the recovery of fine  structure. 
In Tables \ref{Table ImaxCompare} and \ref{Table KodakCompare} we compare representative  $\A\&\B$ methods from the literature with  the best $\B\&\A$ methods identified above (all of them $\B\&1.5\A$): 
\begin{itemize}[leftmargin=*]
  \setlength\itemsep{0em}
\def\labelitemi{--}
    \item The two best performing demosaicking before denoising methods ($\B\&1.5\A$) from on Table~\ref{Table coI} are considered. Namely, RCNN for demosaicking followed by CBM3D (denoted \emph{RCNN+CBM3D}) or nlBayes (\emph{RCNN+nlBayes}) for denoising. 

    \item We also consider a "low-cost" $\B\&1.5\A$ combination using MLRI~\cite{kiku2014minimized} for demosaicking and CBM3D for denoising (\emph{MLRI+CBM3D}).  
\end{itemize}
The considered  $\A\&\B$ methods from the literature are:
\begin{itemize}[leftmargin=*]
  \setlength\itemsep{0em}
\def\labelitemi{--}

\item The  BM3D-CFA  filter was proposed in~\cite{danielyan2009cross} to avoid the checkerboard effects resulting from independently applying BM3D to the color phases of CFA images.
We evaluate 
BM3D-CFA~\cite{danielyan2009cross} followed by Hamilton Adams demosaicking (denoted {\em BM3D+HA}), as well as followed by a state-of-the-art RCNN demosaicking~\cite{tan2017color} ({\em BM3D+RCNN}).
  
  \item 
  The CFA denoising framework of Park \emph{et al.}~\cite{park2009case} effectively compacts the signal energy while the noise is distributed equally in all dimensions by using a color representation from the principal components analysis  of the pixel RGB values in the Kodak dataset and then removes noise in each channel by BM3D. This preprocessing is advantageous  for the Kodak image set, but inadequate for the Imax image set. 
  We evaluate this framework~\cite{park2009case} with BM3D \cite{dabov2006image} followed by the
  RCNN demosaicking~\cite{tan2017color} ({\em Park+RCNN}).


    \item  The PCA-CFA filter proposed in~\cite{zhang2009pca} is a  spatially-adaptive denoising based on 
    principal component analysis (PCA) that exploits the spatial and spectral correlations of CFA images to preserve color edges and details.
    We evaluate PCA-CFA~\cite{zhang2009pca} followed by DLMM demosaiciking~\cite{zhang2005color} ({\em PCA+DLMM}) and RCNN demosaicking~\cite{tan2017color} ({\em PCA+RCNN}).

    \item 
    Finally, as a reference, we include the deep learning based joint denoising and demosaicking  ({\em JCNN}) of~\cite{gharbi2016deep,ehretDEMOSAICK_IPOL2019}. But  its results are only available for noise with $\sigma\le20$ because  the network is not trained beyond that level.
    
\end{itemize}

\begin{table}[t]
\caption{Comparison of the results (CPSNR in dB) between different denoising and demosaicking methods for the {\bf Imax} image set. The best result of each line is in \textcolor{red}{\bf red}, the second best one is in 
\textcolor{green}{\bf green} and the third best one is in \textcolor{blue}{\bf blue}. 
}
\vspace{-0.25cm}
\label{Table ImaxCompare}
\begin{center}
\renewcommand{\arraystretch}{1.25} \addtolength{\tabcolsep}{-5pt} \vskip3mm

\footnotesize

\rowcolors{2}{gray!25}{white}
\begin{tabular}{>{\rowmac}l| >{\rowmac}c >{\rowmac}c >{\rowmac}H >{\rowmac}c >{\rowmac}c >{\rowmac}c |>{\rowmac}c >{\rowmac}c >{\rowmac}c | >{\rowmac}c}
      \rowcolor{gray!50}
\hline 
& \multicolumn{6}{c|}{$\A\&\B$} & \multicolumn{3}{c|}{$\B\&1.5\A$}&\\ 
\hline
      \rowcolor{gray!50}

\setrow{\fontsize{7.5pt}{\baselineskip}\selectfont}      &BM3D&BM3D&Park&Park&PCA&PCA    &RCNN&RCNN& MLRI&\\
           \rowcolor{gray!50}
$\sigma$     &+&+&+&+&+&+    &+&+&+ &JCNN\\
      \rowcolor{gray!50}
\setrow{\fontsize{7.5pt}{\baselineskip}\selectfont}       &HA&RCNN  &HA  &RCNN&DLMM&RCNN&CBM3D&nlBayes&CBM3D&\\\hline

$1$ &34.63& \textcolor{green}{\bf  {38.53}}& 32.74& 35.37& 33.99& 37.52& 38.36& \textcolor{blue}{\bf  {38.42}}&36.52& \textcolor{red}{\bf 38.59}\\
$ 5$ &33.43& \textcolor{red}{\bf 35.62}& 31.57& 32.86& 32.69& 34.87& \textcolor{green}{\bf 35.39}& \textcolor{blue}{\bf  {35.29}}&34.60& 33.48\\
$10$ &31.84& \textcolor{green}{\bf  32.92}& 29.62& 30.06& 30.73& 31.89& \textcolor{blue}{\bf  {32.75}}& 32.59&32.36& \textcolor{red}{\bf 33.09}\\
$20$ &29.22& \textcolor{green}{\bf 29.55}& 26.82& 26.86& 27.57& 27.99& \textcolor{blue}{\bf  {29.41}}& 29.25&29.22& \textcolor{red}{\bf 29.79}\\
$40$ &\textcolor{blue}{\bf  {25.50}}& \textcolor{green}{\bf 25.51}& 23.90& 23.86& 23.50& 23.57& \textcolor{red}{\bf 25.52}& 25.09&25.39& --\\
$60$ &21.55& 21.34&21.78& 21.75& 20.89& 20.89& \textcolor{red}{\bf 22.78}& \textcolor{blue}{\bf  {22.31}}& \textcolor{green}{\bf  {22.63}}& --\\
\hline
Av&28.09& \textcolor{green}{\bf  {28.88}}& 26.45& 26.89& 26.71& 27.53& \textcolor{red}{\bf 28.99}& \textcolor{blue}{\bf  {28.72}}&28.58&--\\
\hline
\end{tabular}
\end{center}
\end{table}

\begin{table}[t]
\caption{Comparison of the results  (CPSNR in dB) between different denoising and demosaicking methods for the {\bf Kodak} image set. The best result of each line is in \textcolor{red}{\bf red}, the second best one is in 
\textcolor{green}{\bf green} and the third best one is in \textcolor{blue}{\bf blue}.
}
\vspace{-0.25cm}
\label{Table KodakCompare}
\begin{center}
\renewcommand{\arraystretch}{1.25} \addtolength{\tabcolsep}{-5pt} \vskip3mm
\footnotesize

\rowcolors{2}{gray!25}{white}
\begin{tabular}{>{\rowmac}l| >{\rowmac}c >{\rowmac}c >{\rowmac}H >{\rowmac}c >{\rowmac}c >{\rowmac}c |>{\rowmac}c >{\rowmac}c >{\rowmac}c | >{\rowmac}c}
      \rowcolor{gray!50}
\hline 
& \multicolumn{6}{c|}{$\A\&\B$} & \multicolumn{3}{c|}{$\B\&1.5\A$}&\\ 
\hline
      \rowcolor{gray!50}

\hline 
\setrow{\fontsize{7.5pt}{\baselineskip}\selectfont
}     & BM3D& BM3D& Park& Park& PCA& PCA    & RCNN& RCNN&  MLRI&\\
           \rowcolor{gray!50}
  $ \sigma$     &+&+&+&+&+&+    &+&+&+ & JCNN\\
       \rowcolor{gray!50}
\setrow{\fontsize{7.5pt}{\baselineskip}\selectfont}       & HA& RCNN  & HA  & RCNN& DLMM& RCNN& CBM3D& nlBayes& CBM3D&\\\hline
$ 1$ &34.70& 40.55& 34.35& 40.36& 38.19& 39.12& \textcolor{green}{\bf  {40.98}}& \textcolor{green}{\bf  {40.98}}& 38.52&\textcolor{red}{\bf 41.15}\\
$ 5$ &32.84& 34.89& 32.54& 34.87& 34.99& 35.42& \textcolor{red}{\bf 36.55}& \textcolor{green}{\bf  {36.42}}&\textcolor{blue}{\bf  {35.71}}& 34.13\\
$10$ &30.34& 30.93& 30.10& 30.85& 31.83& 32.01& \textcolor{red}{\bf 33.36}& \textcolor{blue}{\bf  {33.18}}&32.94& \textcolor{green}{\bf  {33.27}}\\
$20$ &27.59& 27.70& 27.28& 27.42& 28.11& 28.14& \textcolor{red}{\bf 29.98}& \textcolor{blue}{\bf  {29.87}}&29.70 &\textcolor{green}{\bf  {29.95}}\\
$40$ &24.79& 24.78& 24.88&24.88& 24.15& 24.08&\textcolor{red}{\bf 26.71}& \textcolor{blue}{\bf  {26.29}}& \textcolor{green}{\bf  {26.44}}& --\\
$60$ &22.58& 22.55&23.21& 23.19& 21.77& 21.70& \textcolor{red}{\bf 24.42}& \textcolor{blue}{\bf  { 23.93}}&\textcolor{green}{\bf  {24.16}}& --\\
\hline
Av     &27.47& 28.35& 27.41&28.36& 27.96& 28.09& \textcolor{red}{\bf 30.19}&\textcolor{green}{\bf  { 29.93}}& \textcolor{blue}{\bf  {29.64}}&  --\\
\hline
\end{tabular}
\end{center}
\end{table}

From Tables~\ref{Table ImaxCompare} and~\ref{Table KodakCompare} we see that the $\B\&\A$ method RCNN+CBM3D as well as RCNN+nlBayes yield the best results on the Kodak dataset, and the margin with respect to the best $\A\&\B$ method (BM3D+RCNN, i.e. BM3D-CFA~\cite{dabov2007color} with RCNN~\cite{tan2017color}) is quite large: more than 1.5dB on average. 
In Fig.~\ref{fig:compare_kodak} we compare some results obtained on an image from the Kodak dataset. From the upper-left extract we can see that textures are better restored with RCNN+CBM3D and MLRI+CBM3D, while JCNN introduces some defects. From  the extract we see that the $\B\&1.5\A$ methods preserve much more details than BM3D+RCNN, and the result is comparable to JCNN.

On the Imax database RCNN+CBM3D has the highest CPSNRs on high noise levels, by a small gap though. For low noise levels BM3D+RCNN is better, but the difference with RCNN+CBM3D is very small.
The joint denoising-demosaicking network JCNN~\cite{gharbi2016deep} yield the best results  on the Imax dataset for $\sigma\le 20$ (not trained above those levels) yet, the margin with respect to RCNN+CBM3D is again small. 
Overall, by looking at the average CPSNR we can say that the $\B\&1.5\A$ scheme RCNN+CBM3D is indeed much more robust than BM3D+RCNN.



\begin{figure}
\footnotesize
    \centering
    \addtolength{\tabcolsep}{-5pt}
    \begin{tabular}{ccc}
    \overimgx[width=.32\columnwidth, trim={0 50 50 30}, clip]{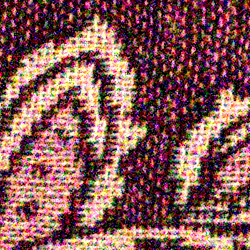}{28.46dB} &
    \overimgx[width=.32\columnwidth, trim={0 50 50 30}, clip]{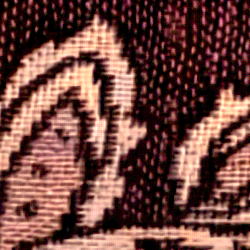}{34.30dB} &
    \overimgx[width=.32\columnwidth, trim={0 50 50 30}, clip]{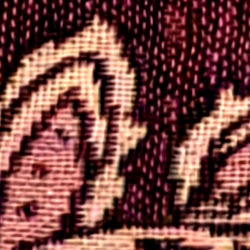}{35.84dB} \\

    \overimgx[width=.32\columnwidth, trim={0 0 0 50}, clip]{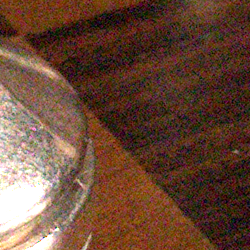}{28.82dB} &
    \overimgx[width=.32\columnwidth, trim={0 0 0 50}, clip]{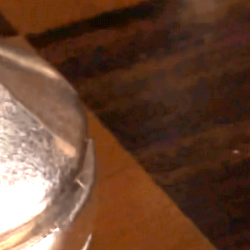}{37.03dB} &
    \overimgx[width=.32\columnwidth, trim={0 0 0 50}, clip]{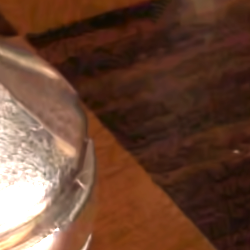}{38.48dB} \\
    noisy demosaicked & $\A\&\B$ & $\B\&1.5\A$
    \end{tabular}
    
    \caption{Details of a real images (enhanced contrast) from the SIDD~\cite{Abdelhamed2018} dataset. From left to right: noisy input (demosaicked), BM3D+RCNN , and RCNN+CBM3D. \vspace{-.5em}}
    \label{fig real images}

\end{figure}

\gf{}{
\paragraph{Evaluation on real images.}

We evaluated on a set of 14 raw images taken from the Small SIDD dataset~\cite{Abdelhamed2018}. For simplicity, the selected images correspond to phones from the same manufacturer. 
We adopted the simple pipeline proposed by the authors, which yields photo finished images that can be compared with the ground truth. 
The considered methods (RCNN+CBM3D, CBM3D+RCNN, and JCNN) were applied at the demosaicking stage (in linear space). 
Before any denoising step ($\A$)  we applied a VST (squared root~\cite{Anscombe1948}), which whitens the noise, and invert it afterwards. 
The noise level was estimated using~\cite{ponomarenko_ipol2013} and provided to the denoising algorithms and JCNN.  

\begin{table}[]
\footnotesize
    \centering
    \begin{tabular}{c|cc|c}
    \hline
      \rowcolor{gray!50}
 mean  $\sigma$  & CBM3D+RCNN & RCNN+CBM3D &  JCNN \\
    \hline
 7.65 &38.19 & \bf 39.64 & 38.54 \\
    \hline
    \end{tabular}
    \caption{Average CPSNR over 14 raw images taken from the Small SIDD dataset~\cite{Abdelhamed2018}. The reported average noise level is scaled to the range 0-255.
    }
    \label{tab:real images}
\end{table}
 
Table~\ref{tab:real images} reports the average CPSNR obtained on these images and the average of the estimated noise levels (after whitening). These values are consistent with the simulated results obtained on the Kodak database (Table~\ref{Table KodakCompare}). 
The result in Fig.~\ref{fig real images}, and the supplementary material, support the case in favor of the $\B\&1.5\A$ schemes (RCNN+CBM3D).
}

\section{Conclusions}
\label{sec:conclusion}

This paper analyzed the advantages and disadvantages of denoising before demosaicking ($\A\&\B$) schemes, versus  demosaicking before denoising ($\B\&\A$), to recover high quality full-color images.
We showed that for the  $\B\&\A$ schemes  a very simple change of the noise parameter of the denoiser $\A$ coped with the structure of demosaicked noise, and led to  efficient denoising \textit{after} demosaicking.
We found that, this allowed  to preserve  fine structures that are often smoothed out by the $\A\&\B$ schemes. 
Our best performing  combination in terms of  quality  and speed is a $\B\&1.5 \A$ scheme, where demosaicking $\B$ is done by a fast algorithm RCNN~\cite{tan2017color} followed by CBM3D denoising $1.5\A$ with noise parameter equal to $1.5\sigma_0$.

 Nevertheless it seems ineluctable to see deep learning win the end game when solutions will be  found to have more compact or more  rapid joint demoisaicking-denoising algorithms.

\medskip

\noindent {\bf Acknowledgments} :   Work partly financed by Office of Naval research  grant N00014-17-1-2552 and  DGA Astrid project  n$^\circ$ ANR-17-ASTR-0013-01.

\newpage

{\small
\bibliographystyle{ieee_fullname}
\bibliography{ridemosaiking}
}

\end{document}